\documentclass[12pt]{iopart}
\usepackage{graphicx}
\usepackage{cite}
\usepackage{CJK}
\usepackage[colorlinks,
linkcolor=blue,
anchorcolor=blue,
citecolor=blue,
urlcolor=blue
]{hyperref}

\begin{document}
\begin{CJK*}{UTF8}{gbsn}
\title[]{Screening effect of plasma flow on the resonant magnetic perturbation penetration in tokamak based on two-fluid model}

\author{Weikang TANG (汤炜康)$^{1,2}$, Qibin LUAN (栾其斌)$^{3*}$, Hongen SUN (孙宏恩)$^{3}$, Lai WEI (魏来)$^{1,2}$, Shuangshuang LU (路爽爽)$^{1}$, Shuai JIANG (姜帅)$^{1}$, Jian XU (徐健)$^{1}$ and Zhengxiong WANG (王正汹)$^{1}$}

\address{$^{1}$Key Laboratory of Materials Modification by Laser, Ion, and Electron Beams (Ministry of Education), School of Physics, Dalian University of Technology, Dalian 116024, People's Republic of China\\
$^{2}$Key Laboratory of Geospace Environment, University of Science and Technology of China, Hefei 230026, People's Republic of China\\
$^{3}$Faculty of Electronic Information and Electrical Engineering, Dalian University of Technology, Dalian 116024, People's Republic of China
}
\ead{luanqb@dlut.edu.cn}
\vspace{10pt}

\begin{abstract}
Numerical simulation on the resonant magnetic perturbation penetration is carried out by the newly-updated initial value code 
MDC (MHD@Dalian Code). Based on a set of two-fluid four-field equations, the bootstrap current, parallel and 
perpendicular transport effects are included appropriately. Taking into account the bootstrap current, 
a mode penetration like phenomenon is found, which is essentially different from the classical tearing mode model. To reveal the influence of the plasma flow on the mode penetration process, $\bf E\times B$ drift flow and diamagnetic drift flow 
are separately applied to compare their effects. Numerical results show that, a sufficiently large diamagnetic
drift flow can drive a strong stabilizing effect on the neoclassical tearing mode. Furthermore,
an oscillation phenomenon of island width is discovered. By analyzing in depth, it is found that, 
this oscillation phenomenon is due to the negative feedback regulation of pressure on the magnetic island. 
This physical mechanism is verified again by key parameter scanning.     
\end{abstract}

\vspace{2pc}

\noindent(Some figures may appear in colour only in the online journal)

\maketitle


\section{Introduction}\label{sec1}
Tearing mode (TM) instability is extensively investigated by researchers in the area of tokamak plasmas in 
the recent decades\cite{FKR,rutherford1973}. The TM is one kind of current driven magnetohydrodynamic (MHD) instabilities commonly 
followed by the magnetic reconnection, which can break up the nested magnetic flux surfaces and generate magnetic 
islands at corresponding resonant surface. These magnetic islands can provide a ``seed'', called seed island, for 
the neoclassical tearing mode (NTM) to grow. NTM, a pressure gradient driven MHD instability, is linearly stable but 
can be destabilized by helical perturbations due to the loss of bootstrap current inside the seed island\cite{seed}. The onset 
of NTM is the principal limitation of the plasma temperature in the core region\cite{Bardoczi2017b}, owing to the  radial ``shortcut'' transport 
in the produced large magnetic island, and one of the main cause of major disruption\cite{Wang2015,Zhangw2019}. 
For the sake of economic feasibility, a high fraction of bootstrap current, up to 80\%--90\%, is required for future advanced tokamak. 
Since NTM is a high beta phenomenon, which is proportional to the bootstrap current fraction, the control and suppression of NTM 
is of great significance for the steady state operation of tokamak devices\cite{LaHaye2006,Sauter_2010,Maraschek2012}. 

Aiming to control the NTM, many research efforts have been dedicated to the resonant magnetic perturbation (RMP)\cite{Hender_1992,yu2009RMP,Buttery_1999,Lanctot_2016,Lu_2020,Logan_2021,rmpreview}. 
RMP has been found to drive additional effects on magnetic islands in tokamak plasmas. Specifically, the RMP can produce an 
electromagnetic torque at the corresponding resonant surface. Once the electromagnetic torque is sufficiently large to balance the plasma viscosity 
and inertia torque, the magnetic islands would be compulsorily aligned with the RMP in an identical frequency, called locked mode (LM)\cite{Nave_1990}. 
For a static RMP, it can be used to test the maximum tolerance for the residual error field, resulting from the asymmetry of the tokamak device\cite{Wanghuihui}. 
As for the dynamic RMP, it can be utilized to unlock the magnetic island and maintain a stable toroidal and poloidal rotation\cite{Wolf_2005}. 
Lately, experimental and numerical results show that the synergetic application of RMP and electron cyclotron current drive (ECCD) is 
a promising and effective method to control the NTM\cite{Choi2018,Tang_2020,Nelson_2020}. The RMP can be used as an auxiliary method to lock and locate the phase of 
the NTM, and then to enhance the accuracy and effectiveness of the ECCD. 

In addition to above application, even if the rotating plasma is originally stable to the NTM, the RMP can drive magnetic 
reconnection and generate magnetic island at the resonant surface, called mode penetration\cite{Fitzpatrick_1993,Yu_2008}. Mode penetration has raised much concern, since 
its threshold is directly related to the onset of TM/NTM. Based on single-fluid theory, considerable studies have been made in predicting 
and explaining the threshold of mode penetration for different tokamak devices\cite{Buttery_1999,Buttery_2000,Haye_1992,Wang_2014,Wang_2018} and parameter regimes\cite{cole07prl,wangjl2015,beidler_2018,Zhang_2021}. However, considering the plasma rotation 
playing a significant role in the screening process of RMP\cite{Becoulet2012}, the two-fluid model, retaining the electron diamagnetic drift as well as 
the ${\bf E\times B}$ flow, is more suitable to account for more complex physics. Recently, in the frame of the two-fluid drift-MHD theory, 
plenty of researches were carried out to investigate the interaction of the RMP and magnetic islands in tokamak plasmas\cite{Yu2018,Fitz18popa,Fitz18popb}. 
Using a two-fluid model, Hu {\it et al} found that the two-fluid effects can give significant modifications to the scaling law of mode penetration for different 
plasma parameters. Besides, the enduring mystery, non-zero penetration threshold at zero plasma natural frequency, is 
explained by the small magnetic island width when penetrated\cite{Hu_2020,De_Bock_2008}. In the recent investigation, numerical results show that the 2/1 NTM can be suppressed by the 4/2 RMPs with moderate amplitudes, if the bi-normal fluid rotation frequency is in the ion diamagnetic drift direction or sufficiently large\cite{Yu_2021}.

Motivated by the above reasons, based on the two-fluid model, the screening effects of two component of plasma flows on the mode penetration are investigated in this work. Taking into account 
the bootstrap current, a mode penetration like process is found by numerical simulation. Furthermore, it is found that the diamagnetic 
drift flow has a stabilizing effect on the magnetic islands. An oscillation phenomenon of island width is discovered in high Lundquist number $S$ 
and high transport scenario. 

The rest of this paper is organized as follows. In section \ref{sec2}, the modeling equations used in this 
work are introduced. In section \ref{sec3}, numerical results and physical discussions are presented. Finally, 
the paper is summarized and conclusions are drawn in section \ref{sec4}.

\section{Physical model}\label{sec2}
The initial value code MDC (MHD@Dalian Code)\cite{Wei2016,Wang2017,Liu2018,Ye2019,JIANG_2022,Han_2017,WW_2018} is upgraded to the two fluid version based on a set of four-field MHD equations\cite{hazeltine1985}. 
Taking into account the nonlinear evolution of the vorticity $U$, the poloidal magnetic flux $\psi$, the plasma pressure $p$ and the parallel ion velocity $v$, 
the normalized equations in the cylindrical geometry ($r$, $\theta$, $z$) can be written as
\begin{equation}
\frac{\partial U}{\partial t} = [U,\phi+\delta\tau p] + \nabla_\parallel j + \nu\nabla^2_\perp U + \delta\tau[\nabla_\perp p,\nabla_\perp U],\label{eq1}
\end{equation}
\begin{equation}
\frac{\partial \psi}{\partial t} = -\nabla_{\parallel}\phi + \delta\nabla_{\parallel}p - \eta(j - j_{\rm b}),\label{eq2}
\end{equation}
\begin{equation}
\frac{\partial p}{\partial t} = [p,\phi] + 2\beta_{\rm e}\delta\nabla_\parallel j -\beta_{\rm e}\nabla_\parallel v + \chi_\parallel\nabla^2_\parallel p + \chi_\perp\nabla^2_\perp p,\label{eq3}
\end{equation}
\begin{equation}
\frac{\partial v}{\partial t} = [v,\phi] - \frac{1}{2}(1+\tau)\nabla_\parallel p + \mu\nabla^2_\perp v,\label{eq4}
\end{equation}
where $\phi$ and $j$ are, respectively, the electric potential and plasma current density along the axial direction, obtained by the following formulas 
$U=\nabla^2_\perp \phi$ and $j=-\nabla^2_\perp \psi$. The equation (\ref{eq1}) (vorticity equation) is the perpendicular component (taking $e_z\cdot\nabla\times$) 
of the equation of motion, where $\nu$ is the viscosity and $\tau=T_{\rm i}/T_{\rm e}$ is the ratio of ion to electron temperature. 
$\delta=(2\Omega_{\rm i}\tau_{\rm a})^{-1}$ is a gyroradius parameter, where $\Omega_{\rm i}=eB_0/m_{\rm i}$ is a constant measure of the ion gyrofrequency 
and $\tau_{\rm a}=\sqrt{\mu_0\rho}a/B_0$ is Alfv\'{e}n time. Neglecting the electron inertia and Hall effect, the equation (\ref{eq2}) is obtained by combining 
the generalized Ohm's law and Faraday's law of electromagnetic induction, where $\eta$ is the resistivity. 
$j_{\rm b}=-(A\sqrt\varepsilon/B_\theta) p'(r)$ is the bootstrap current\cite{bs}, where $\varepsilon = a/R_0$ is the 
inverse aspect-ratio, and $B_\theta$ is the poloidal magnetic field. $A$ is a constant that can be calculated by a given bootstrap current fraction $f_{\rm b} = \int_0^a j_{\rm b}r{\rm d}r/\int_0^a jr{\rm d}r$. 
Since isothermal assumption is made here, the evolution of pressure is mainly determined by the particle conservation law. 
It is basically a transport equation, considering the convective term, parallel and perpendicular heat transport. By including the effect of resistive diffusion and the parallel ion flow, 
the equation (\ref{eq3}) is the final energy transport equation for two-fluid plasma. $\beta_{\rm e}=2\mu_0n_0T_{\rm e}/B_0^2$ is the electron plasma beta at the location of magnetic axis, 
where $T_{\rm e}$ is the constant electron temperature. $\chi_\parallel$ and $\chi_\perp$ are the parallel and perpendicular transport coefficients, respectively. 
The equation (\ref{eq4}) is the parallel component of the equation of motion by taking the dot product of the equation with $B$, where  
$\mu$ is the diffusion coefficient of parallel ion velocity. This model can reduce to the high-beta reduced MHD equations of Strauss\cite{strauss_highbMHD}, 
by giving the limit $\delta\rightarrow 0,\beta_{\rm e}\rightarrow 0$. New physics appears by introducing the two factors $\delta$, 
measuring the finite Larmor radius (FLR) effects, and $\beta_{\rm e}$, measuring the compressibility. 
If one has $\delta\rightarrow 0$ and $T_{\rm i}=T_{\rm e}$, but nonzero $\beta_{\rm e}$, the model reduces to the compressible 
reduced MHD (CRMHD) equations\cite{CRMHD}. 

The length $l$, time $t$, and velocity $v$ 
are normalized by the minor radius $a$, Alfv\'{e}n time $\tau_{\rm a}$ and Alfv\'{e}n speed $v_{\rm a}=B_0/\sqrt{\mu_0\rho}$, respectively.  
The poloidal magnetic flux $\psi$, electric potential $\phi$ and plasma pressure $p$ are normalized by $aB_0$, $aB_0v_{\rm a}$ and the pressure 
at the magnetic axis, respectively. The normalization of the diffusion coefficients is as follows, $\eta$, $\nu$, $\mu$, $\chi_\parallel$, $\chi_\perp$ are normalized by $\mu_0a^2/\tau_{\rm a}$, $a^2/\tau_{\rm a}$, $a^2/\tau_{\rm a}$, $a^2/\tau_{\rm a}$, $a^2/\tau_{\rm a}$, respectively.
The Poisson brackets in equations~(\ref{eq1})--(\ref{eq4}) are defined as
\begin{equation} 
[f,g]=\nabla f\times\nabla g\cdot \hat{z}=\frac{1}{r}(\frac{\partial f}{\partial r}\frac{\partial g}{\partial \theta}-\frac{\partial g}{\partial r}
\frac{\partial f}{\partial \theta}).\label{eq5}
\end{equation}
Each variable $f(r,\theta,z,t)$ in equations~(\ref{eq1})--(\ref{eq4}) can be written in the form $f = f_0(r) + \widetilde{f}(r,\theta,z,t)$ with $f_0$ and $\widetilde{f}$ 
being the time-independent initial profile and the time-dependent perturbation, respectively. By applying the periodic boundary conditions in the poloidal and axial
directions, the perturbed fields can be Fourier-transformed as
\begin{equation}
\widetilde{f}(r,\theta,z,t)=\frac{1}{2}\sum_{m,n}\widetilde{f}_{m,n}(r,t){\rm e}^{{\rm i}m\theta-{\rm i}nz/R_0},\label{eq6}
\end{equation}
with $R_0$ being the major radius of the tokamak, $m$ and $n$ being the poloidal and toroidal mode number, respectively.

The effect of RMP with $m/n$ is taken into account by the boundary condition
\begin{equation} 
    \widetilde{\psi}_{m,n}(r=1)=\psi_{\rm a}(t){\rm e}^{{\rm i}m\theta-{\rm i}nz/R_0}.
\end{equation}
In this way, the perturbed radial magnetic field at plasma boundary over toroidal magnetic field could be calculated by 
$\delta B_r/B_0 = am\psi_{\rm a}$.  
It should be pointed out that, in a real tokamak, the toroidal rotation is prevailing and much stronger than the poloidal
one, whereas only the poloidal rotation is considered in this work. Considering the fact that the electromagnetic
force exerted in the poloidal direction is $(n/m)(r_{\rm s}/R)$ times smaller than that in toroidal direction, where $r_{\rm s}$ is the radial location of the resonant surface, and the speed in
toroidal direction should be $(m/n)(R/r_{\rm s})$ times larger than the poloidal one for having an equivalent rotation frequency,
the locking threshold in the toroidal direction can, therefore, be estimated by multiplying such a factor $[(m/n)(R/r_{\rm s})]^2$.

Given the initial profiles of $\phi_0$, $\psi_0$, $p_0$ and $v_0$, equations~(\ref{eq1})--(\ref{eq4}) can be solved simultaneously by our code MDC. 
The two-step predictor-corrector method is applied in the time advancement. The finite difference method is used in the radial 
direction and the pseudo-spectral method is employed for the poloidal and the toroidal directions ($\theta,\zeta=-z/R_0$).

\section{Simulation results}\label{sec3}
\subsection{Numerical set-up}
\begin{figure*}[htb!]
\centering
\includegraphics[width=1.\textwidth]{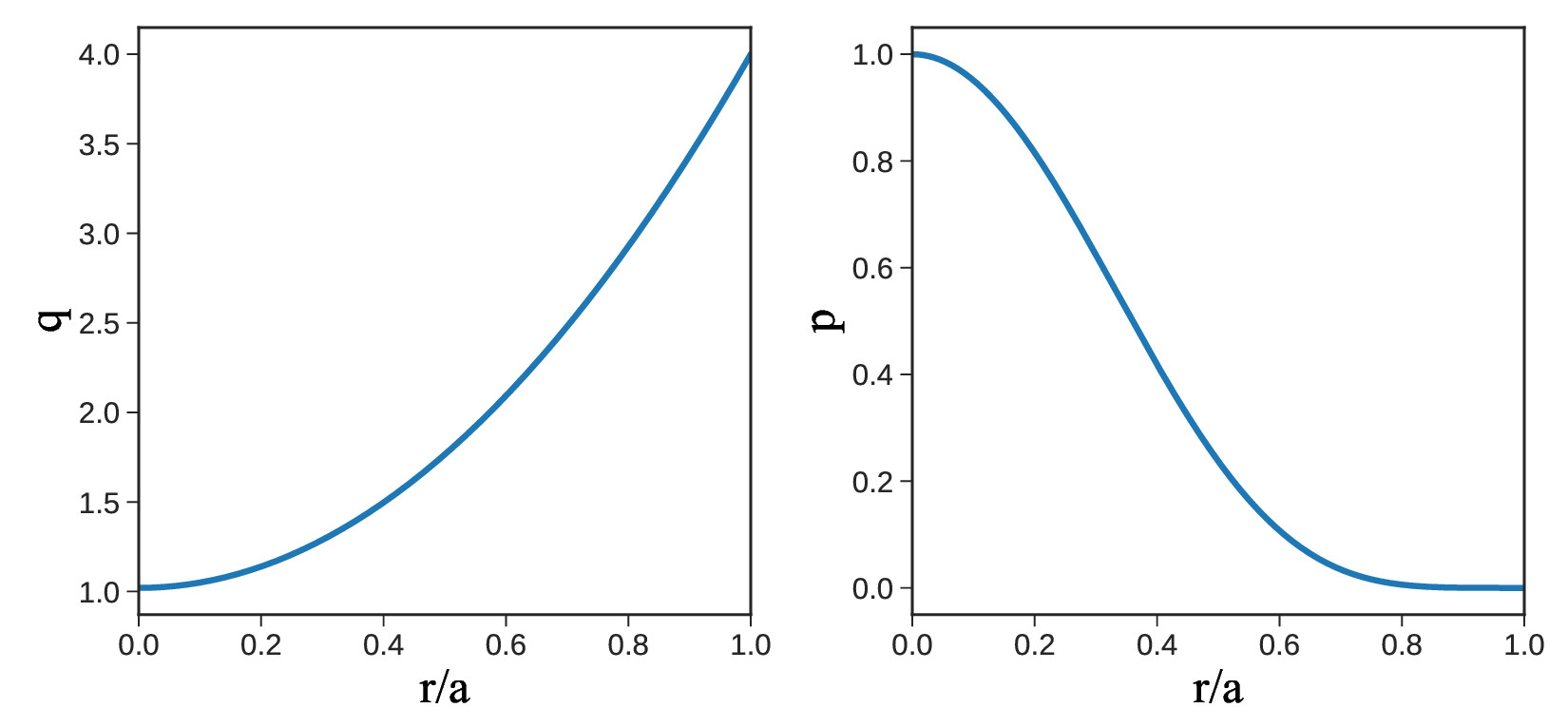}
\caption{Safety factor $q$ and pressure $p$ profiles adopted in this work.}
\label{fig1}
\end{figure*}

Considering a low density ohmically heated tokamak discharge with electron density $n_{\rm e}\approx2\times10^{19}\rm \ m^{-3}$, 
toroidal magnetic field $B_0=2$\ T and $\varepsilon=0.25$, 
this will lead to the Alfv\'{e}n speed $v_{\rm a}\approx 6.9\times10^6\rm\ m/s$ and Alfv\'{e}n time $\tau_{\rm a}\approx7.24\times10^{-8}\rm \ s$.
The corresponding Alfv\'{e}n frequency is $\omega_{\rm a}\approx 1.38\times10^7\rm\ Hz$.
Otherwise stated, other plasma parameters are set as follows, $\tau=1$, $\beta_{\rm e}=0.01$, $\eta=10^{-6}$, $\nu=10^{-7}$, $\mu=10^{-6}$, $\chi_{\parallel}=10$ and $\chi_{\perp}=5\times10^{-6}$. In the experimental condition, $\chi_{\parallel}/\chi_{\perp}$ can be $10^8-10^{10}$. However, in our simulation, this value is a little lower than the realistic one due to computational limitation. The radial mesh number is set as $N_{\rm r}=2048$. The typical time step $\Delta t$ in the simulations is chosen as 0.001. However, $\Delta t$ varies from different cases due to numerical stability. In this work, the nonlinear
simulations only include single helicity perturbations with higher harmonics ($m/n=3/2$, 3 $\leq$ $m$ $\leq$ 18), in addition to 
the changes in the equilibrium quantities ($m/n=0/0$ component). To simulate the mode penetration process,  
a linearly stable equilibrium safety factor $q$ profile and the normalized plasma pressure $p$ 
profile $p_0(r) = (1 - r^2)^5$ are given in figure \ref{fig1}, with the $q=3/2$ resonant surface located at $r=0.402$.
\begin{figure*}[htb!]
\centering
\includegraphics[width=1.\textwidth]{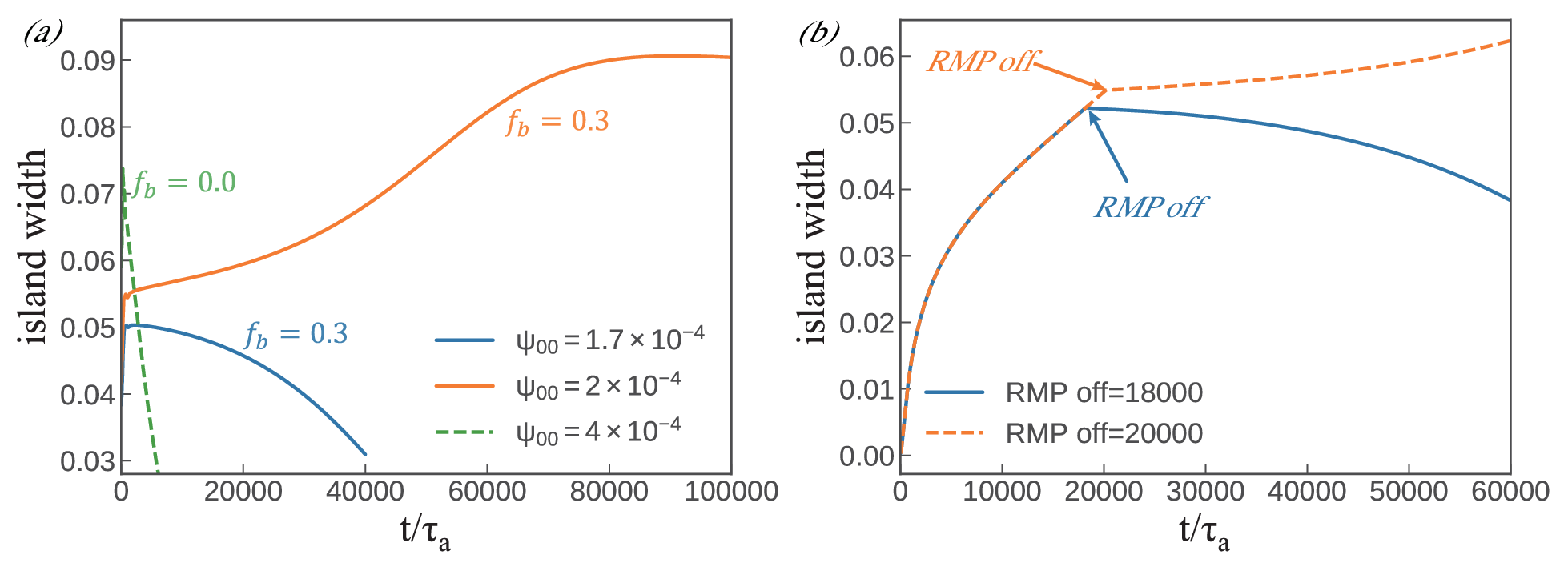}
\caption{(a) Nonlinear evolution of the island width for different magnitudes 
of seed island. The solid traces are for bootstrap current fraction $f_{\rm b}=0.3$ 
and the dashed trace is for $f_{\rm b}=0$. (b) Nonlinear evolution of island
width for different turn-off time of RMP with $f_{\rm b}=0.3$ and $\delta B_r/B_0=3.75\times10^{-5}$.}
\label{fig2}
\end{figure*}

\subsection{Basic verification}
To begin with, the role of seed island width on the onset of NTM is verified to ensure the neoclassical current effect
implemented properly. Under real experimental conditions, the seed island comes from a variety of sources, and it may 
come from different MHD modes, so the seed island width is considered by the initial magnetic perturbation in an arbitrary form of 
$\widetilde{\psi}_{t=0}=\psi_{00}(1-r)^2$. The nonlinear evolution of magnetic island width for different initial 
magnetic perturbations are presented in figure \ref{fig2} (a). The solid traces are for bootstrap 
current fraction $f_{\rm b}=0.3$ (NTM) and dotted one for $f_{\rm b}=0.0$ (TM). For the classical TM, even if a very large seed  
island width is given, the island width still fades with time, illustrating that the TM in this $q$ profile is 
linearly stable. Taking the bootstrap current into consideration, it is seen that there is a threshold for the 
mode to grow, manifesting that the nonlinearly unstable NTM is triggered for a larger seed island width. 
In experiments, the RMP coils are commonly used to seed a magnetic island. Then the onset of NTM by RMP is 
tested ($\widetilde{\psi}_{t=0}=0$). The RMP is turned on from the very beginning with the amplitude of $\delta B_r/B_0=3.75\times10^{-5}$. 
In figure \ref{fig2} (b), in the presence of RMP, the nonlinear evolution of island width for $f_{\rm b}=0.3$ is shown. 
After applying the RMP, the originally stable TM can be driven unstable, as the island width grows even  
without seed island. Then the RMP is turned off at different time when island width grows to a moderate magnitude,
to testify if it is the driven reconnection or the onset of NTM. It turns out that, if the RMP is turned off before the 
island width is sufficiently large, the magnetic island still could not grow spontaneously. It confirms again  
the existence of the critical island width for triggering the NTM, which is consistent with the theory.

\subsection{Effects of electric drift flow}
\begin{figure*}[htb!]
\centering
\includegraphics[width=1.\textwidth]{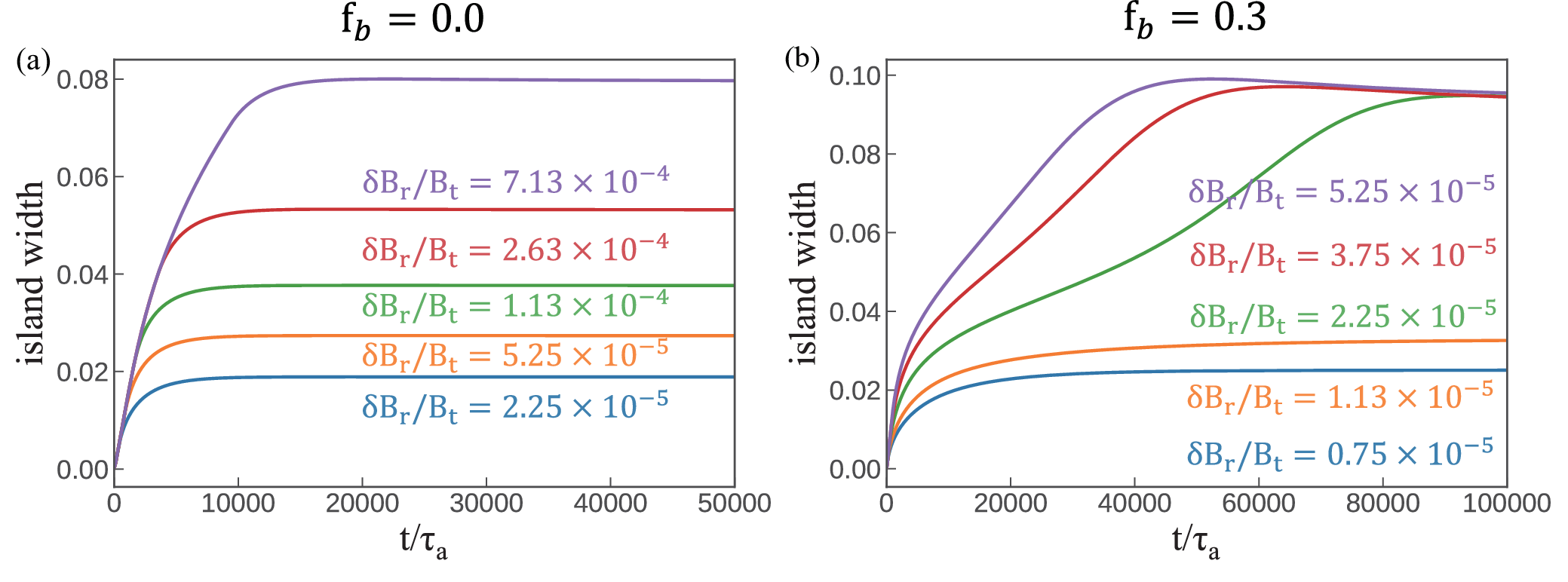}
\caption{Temporal evolutions of island width for different RMP amplitudes for bootstrap current
fraction $f_{\rm b}=0$ (a) and $f_{\rm b}=0.3$ (b) without plasma rotation.}
\label{fig3}
\end{figure*}

Based on the above results, effects of different RMP amplitudes are then scanned for $f_{\rm b}=0.0$ and $f_{\rm b}=0.3$ with zero 
rotation (electric drift $\omega_{\rm E}$ and diamagnetic drift $\omega_{\rm dia}$ are both zero). For $f_{\rm b}=0.0$
(figure \ref{fig3} (a)), the saturated island width is positively related to the RMP amplitude,
a typical driven reconnection. However, for $f_{\rm b}=0.3$ (figure \ref{fig3} (b)), a mode penetration like phenomenon 
can be observed. If the RMP amplitude is relatively small, the magnetic island is going to saturate at a very
small magnitude. Once the RMP amplitude is sufficiently large, the final island width would be very large and keep 
almost the same even further increasing the RMP amplitude. On the other hand, this phenomenon is different from the so-called mode penetration, and we will demonstrate it next.
\begin{figure*}[htb!]
\centering
\includegraphics[width=1.\textwidth]{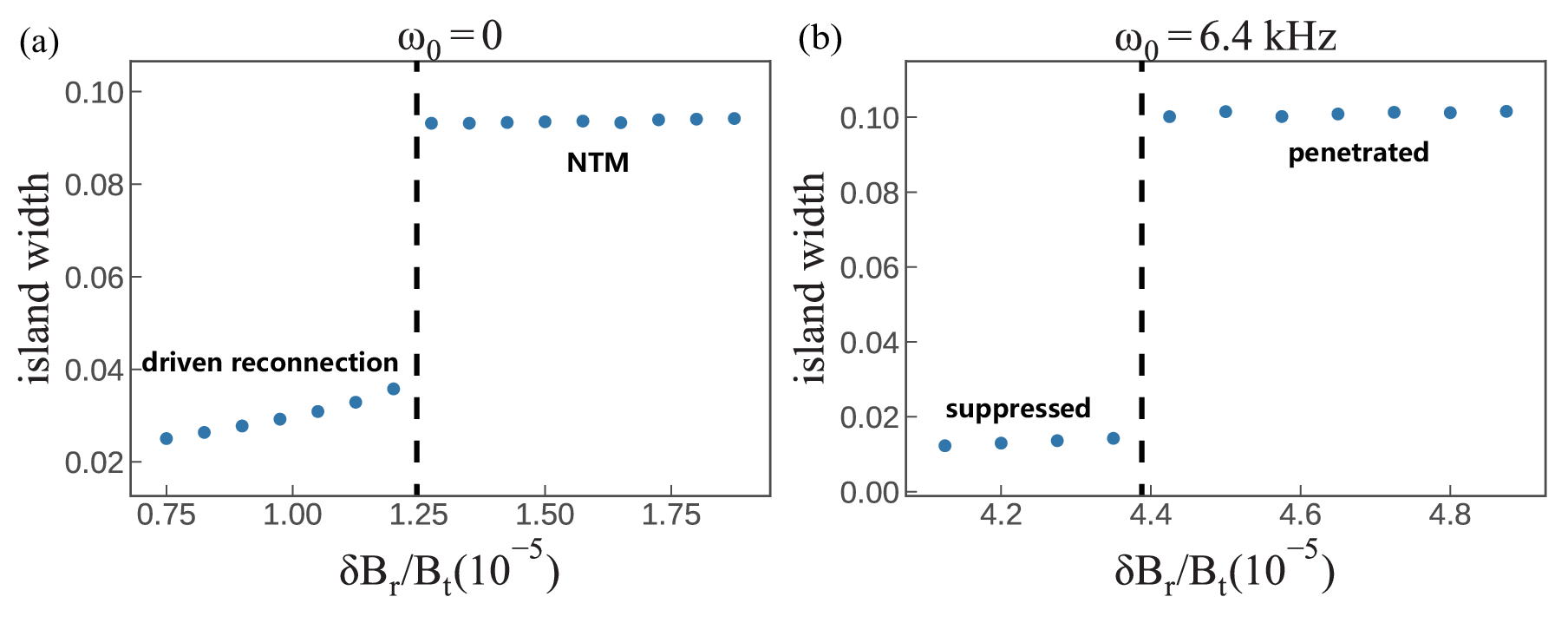}
\caption{Island width versus the RMP amplitude for natural frequency $\omega_0=0$ (a)
and $\omega_0=6.4$ kHz (b). The bootstrap current fraction $f_{\rm b}$ is set as 0.3.}
\label{fig4}
\end{figure*}
\begin{figure*}[htb!]
\centering
\includegraphics[width=1.\textwidth]{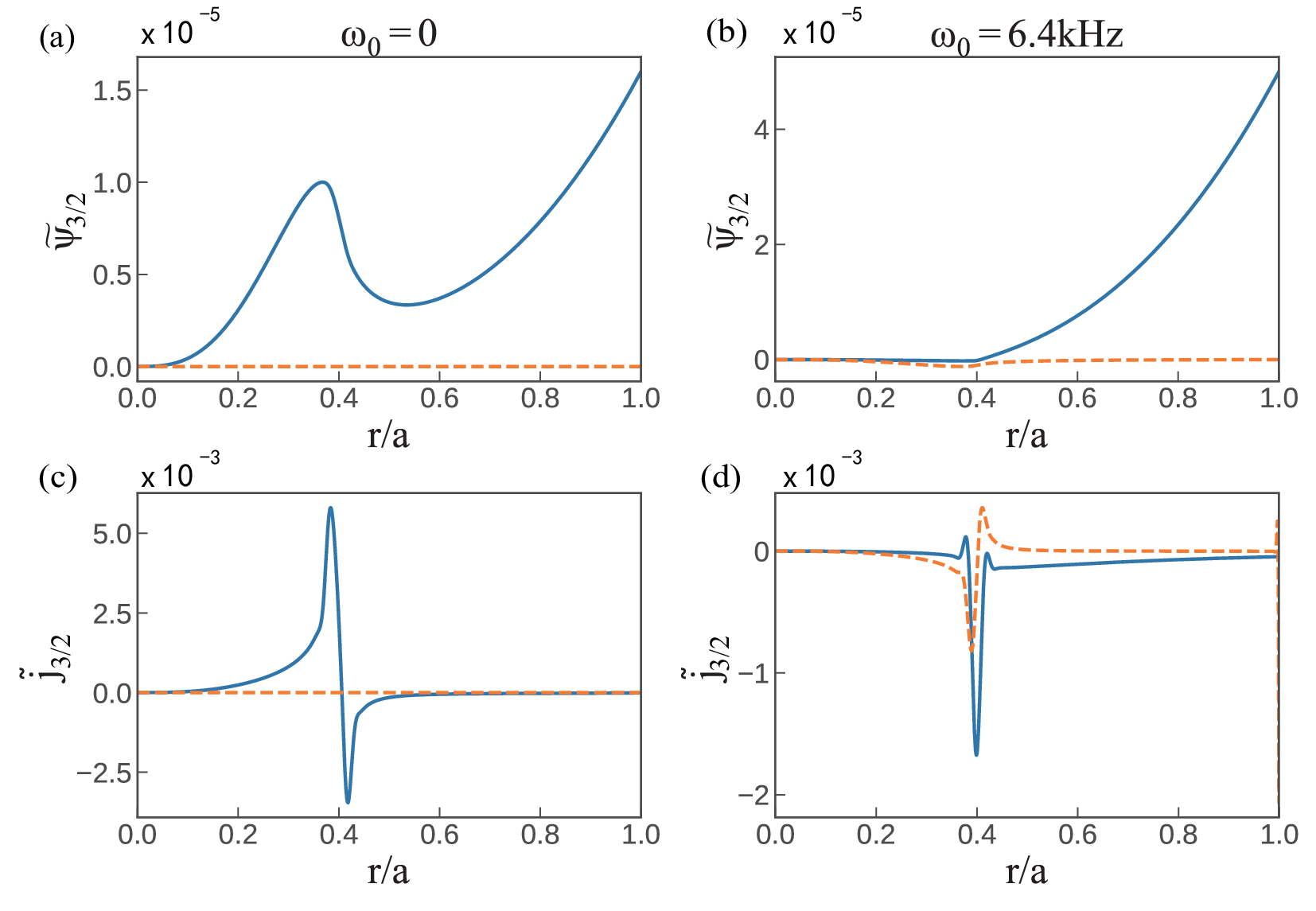}
\caption{The typical eigenmode structure of poloidal magnetic flux and
current density for $m/n=3/2$ component in the regime below the jump of island width.
For natural frequency $\omega_0=0$ (a) and (c), the magnetic perturbation in the core 
region is excited by the RMP. For $\omega_0=6.4$ kHz (b) and (d), the RMP is kept out
from the resonant surface. The solid traces are for the real part and dashed traces
for the imaginary part.}
\label{fig5}
\end{figure*}

Corresponding to figure \ref{fig3} (b), the saturated magnetic island width is shown as a function of RMP 
amplitude in figure \ref{fig4} (a). As the RMP amplitude increases, the saturated island width increases slowly 
first. Once the RMP amplitude exceeds a threshold, there is a jump of the saturated island width. This process can be
divided into two parts, i.e. the driven reconnection phase for smaller RMPs and the onset of NTM for large RMPs. 
Next, including the effect of equilibrium ${\bf E\times B}$ flow, scan over the RMP amplitude is performed again. The           
equilibrium ${\bf E\times B}$ flow is considered by a poloidal momentum source $\Omega_{\rm s}(r)=1/r*{\rm d}\phi_0/{\rm d}r$. 
Here, $\delta$ is set to be zero, so only the ${\bf E\times B}$ flow is considered. For a 6.4 kHz plasma rotation, 
the saturated island width versus the RMP amplitude is presented in figure \ref{fig4} (b). This is a typical 
mode penetration case, as the RMP amplitude increases above a threshold, the island width boosts to a large magnitude.  
There is one main distinction between figure \ref{fig4} (a) and (b), even though they look similar.
That is, below the threshold, the island width barely changes for the case with plasma rotation but slowly increases
for the case without rotation. The increase regime for the case without rotation is due to the forced reconnection 
driven by the externally applied RMP. However, the regime below the threshold for the case with rotation is 
due to the screening effect of the plasma flow. The corresponding eigenmode structures of perturbed poloidal 
magnetic flux and current density for the case without/with plasma rotation are plotted in figure \ref{fig5}. It can be clearly observed that the RMP is screened inside the resonant surface
for the case with rotation, from the fact that the magnetic perturbation is nearly zero for $r<r_{\rm s}$.
In addition, a strong shielding current is formed at the resonant surface, which is consistent with the results on TEXTOR tokamak\cite{kiku2006}. These characteristics are the same as the small locked island phenomenon in \cite{Huqm2012,Tang2019}, suggesting that the small locked island is the complete suppression of magnetic island by the RMP. Besides, recent study also shows similar feasibility of magnetic island suppression by RMP at moderate amplitude\cite{Nies_2022}. For the case without rotation, on the other hand, the $m/n$ = 3/2 component of magnetic perturbation is induced and kept by the 3/2 RMP at the boundary, indicating that mode penetration has already taken place.   

All in all, the above simulation results suggest that, without plasma flow, the penetration threshold for the RMP is zero. Considering the effect of bootstrap current, however, the onset of NTM can lead to a mode penetration like phenomenon,
a driven connection regime plus a NTM regime. As a result, it seems that there is a finite threshold for mode penetration if not carefully distinguished.   

\subsection{Effects of diamagnetic drift flow}
In this subsection, effects of diamagnetic drift flow on the RMP penetration are numerically studied 
in comparison with the ${\bf E\times B}$ electric drift flow. The equilibrium ${\bf E\times B}$ flow velocity can 
be directly obtained by $v_{\rm E0}=\partial\phi_0/\partial r$, and the equilibrium electron diamagnetic flow 
velocity can be calculated by $v_{\rm dia0}=-\delta\partial p_0/\partial r$, where the subscript 0 is for
equilibrium quantities. Therefore, the natural plasma rotation could be obtained by the sum of these two effects.
By changing the value of $\delta$, different amplitudes of diamagnetic drift flow can be implemented.

\begin{figure*}[htb!]
\centering
\includegraphics[width=.5\textwidth]{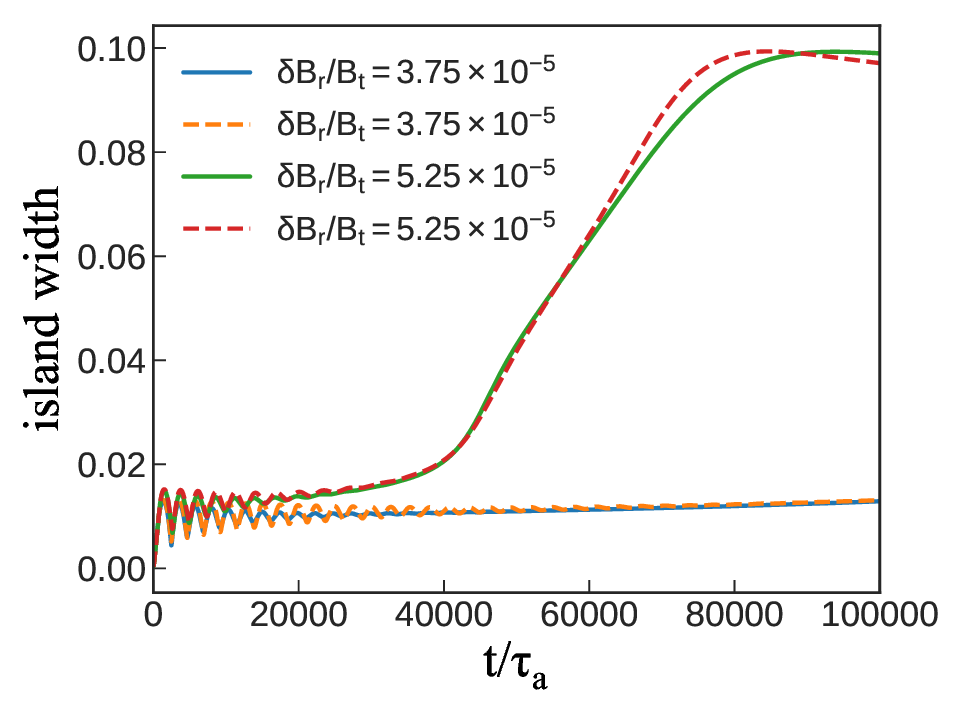}
\caption{Nonlinear evolution of the island width for $\omega_0=\omega_{\rm E0}$=6.4 kHz (solid)
and $\omega_0=\omega_{\rm dia0}$=6.4 kHz (dashed).}
\label{fig6}
\end{figure*}
\begin{figure*}[htb!]
\centering
\includegraphics[width=1.\textwidth]{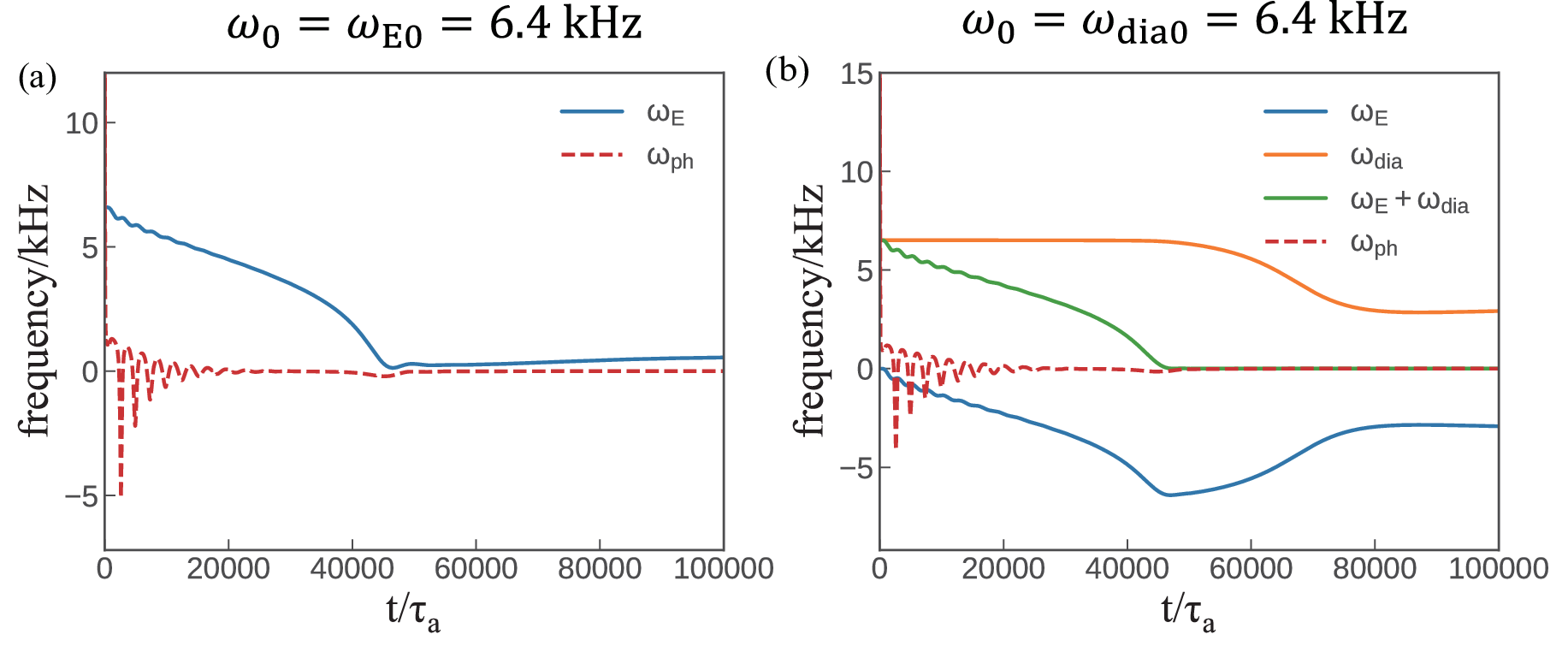}
\caption{Temporal evolutions of the phase frequencies $\omega_{\rm ph}$ and
different flow frequency $\omega_{\rm E}$, $\omega_{\rm dia}$ and
$\omega_{\rm tot}=\omega_{\rm E}+\omega_{\rm dia}$. The $\omega_{\rm ph}$
is calculated by the partial derivative of the phase of poloidal flux $\Phi_\psi$
with respect to time. $\omega_{\rm E}$ and $\omega_{\rm dia}$ are the angular
frequencies of electric drift and diamagnetic drift flow, respectively.}
\label{fig7}
\end{figure*}
\begin{figure*}
\centering
\includegraphics[width=.5\textwidth]{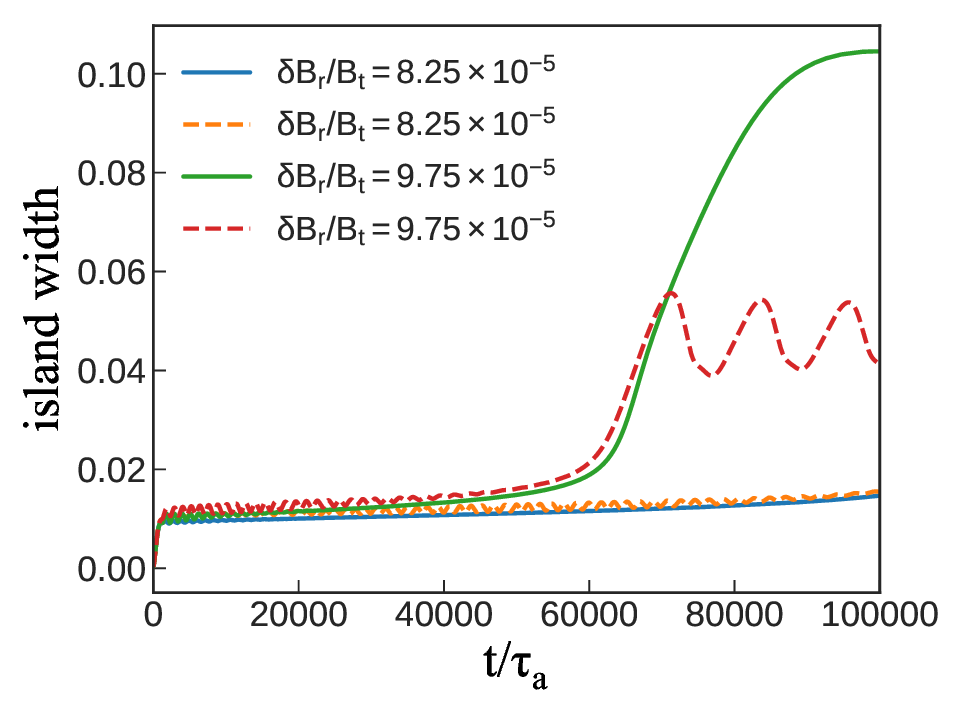}
\caption{Nonlinear evolution of the island width for $\omega_0=\omega_{\rm E0}$=12.8 kHz (solid)
and $\omega_0=\omega_{\rm dia0}$=12.8 kHz (dashed).}
\label{fig8}
\end{figure*}
\begin{figure*}
\centering
\includegraphics[width=1.\textwidth]{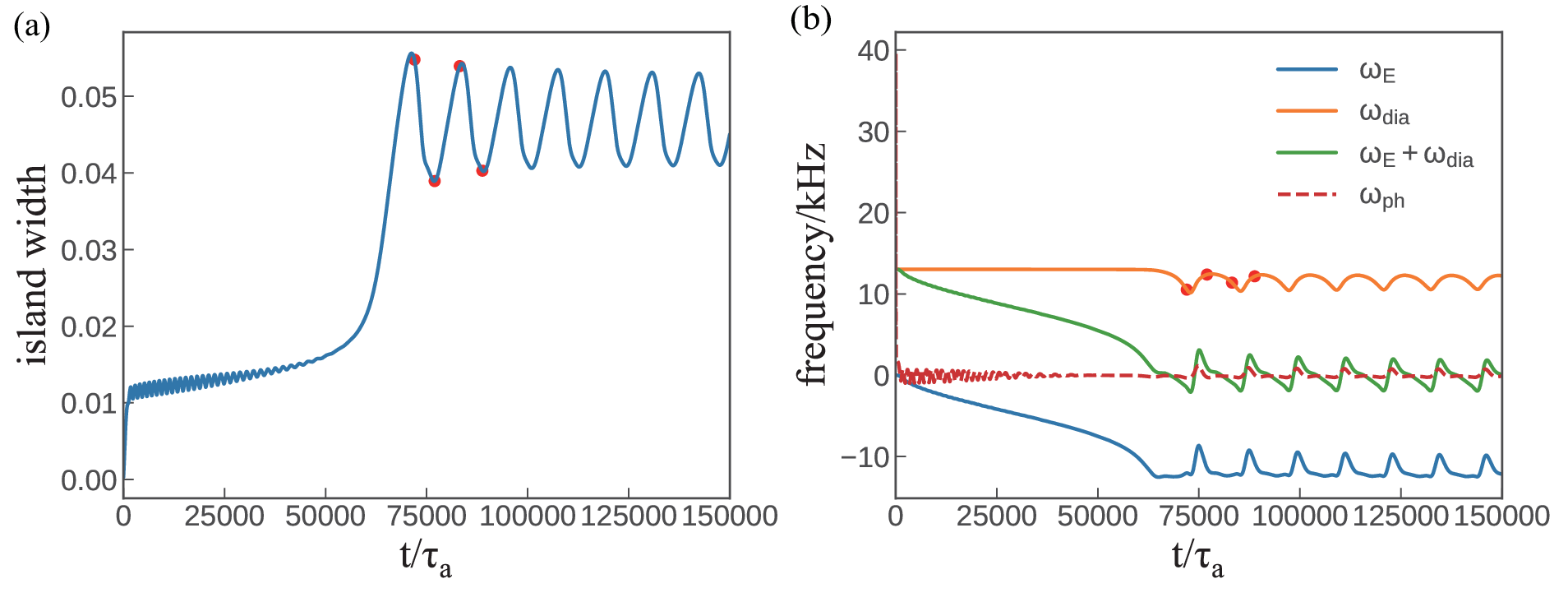}
\caption{Temporal evolution of the island width and various frequencies for the
oscillated case ($\omega_0=\omega_{\rm dia0}$=12.8 kHz, $\delta B_r/B_0=9.75\times 10^{-5}$)
in figure \ref{fig8}. Oscillation of $\omega_{\rm dia}$ is observed after mode penetration.
Four time points $t$=73000, $t$=77000, $t$=83200 and $t$=88800 are marked by the red circles.}
\label{fig9}
\end{figure*}

As shown in figure \ref{fig6}, the screening effects of the two types of flow on the RMP penetration are compared.
The solid lines are for cases with a equilibrium ${\bf E\times B}$ flow frequency $\omega_{\rm E0}$=6.4 kHz and
diamagnetic flow frequency $\omega_{\rm dia0}$=0, while dotted lines for $\omega_{\rm dia0}$=6.4 kHz and
$\omega_{\rm E0}$=0. Thus, for both cases the natural frequency is $\omega_0=\omega_{\rm E0}+\omega_{\rm dia0}$=6.4
kHz. It makes no difference on the penetration threshold, no matter what kind of the flow it is, which means that the
penetration threshold only depends on the rotation difference between the resonant surface and the RMP.
Corresponding to the two penetrated cases in figure \ref{fig6}, the temporal evolutions 
of the flow frequencies $\omega_{\rm E}$, $\omega_{\rm dia}$, $\omega_{\rm E}$+$\omega_{\rm dia}$
and the frequency of the phase of the island $\omega_{\rm ph}$ are illustrated in figure \ref{fig7}.
It shows a good agreement for the flow frequency and the phase frequency after mode penetration, indicating that
the magnetic island and the flow are coupled in accordance with the frozen-in theorem. For the case ($\omega_{\rm E0}$=6.4 kHz,
$\omega_{\rm dia0}$=0), the $\omega_{\rm E}$ decreases with time and drops to zero at the moment penetration occurs.
For the case ($\omega_{\rm E0}$=0, $\omega_{\rm dia0}$=6.4 kHz), since the $\omega_{\rm dia}$ is a rigid flow effect which is mainly
proportional to the pressure gradient, it does not change at the beginning but starts to
decrease as the magnetic island grows, where the pressure gradient is flattened. In consequence, the $\omega_{\rm E}$
will rotate in the opposite direction to cancel the diamagnetic flow.

\begin{figure*}
\centering
\includegraphics[width=1.\textwidth]{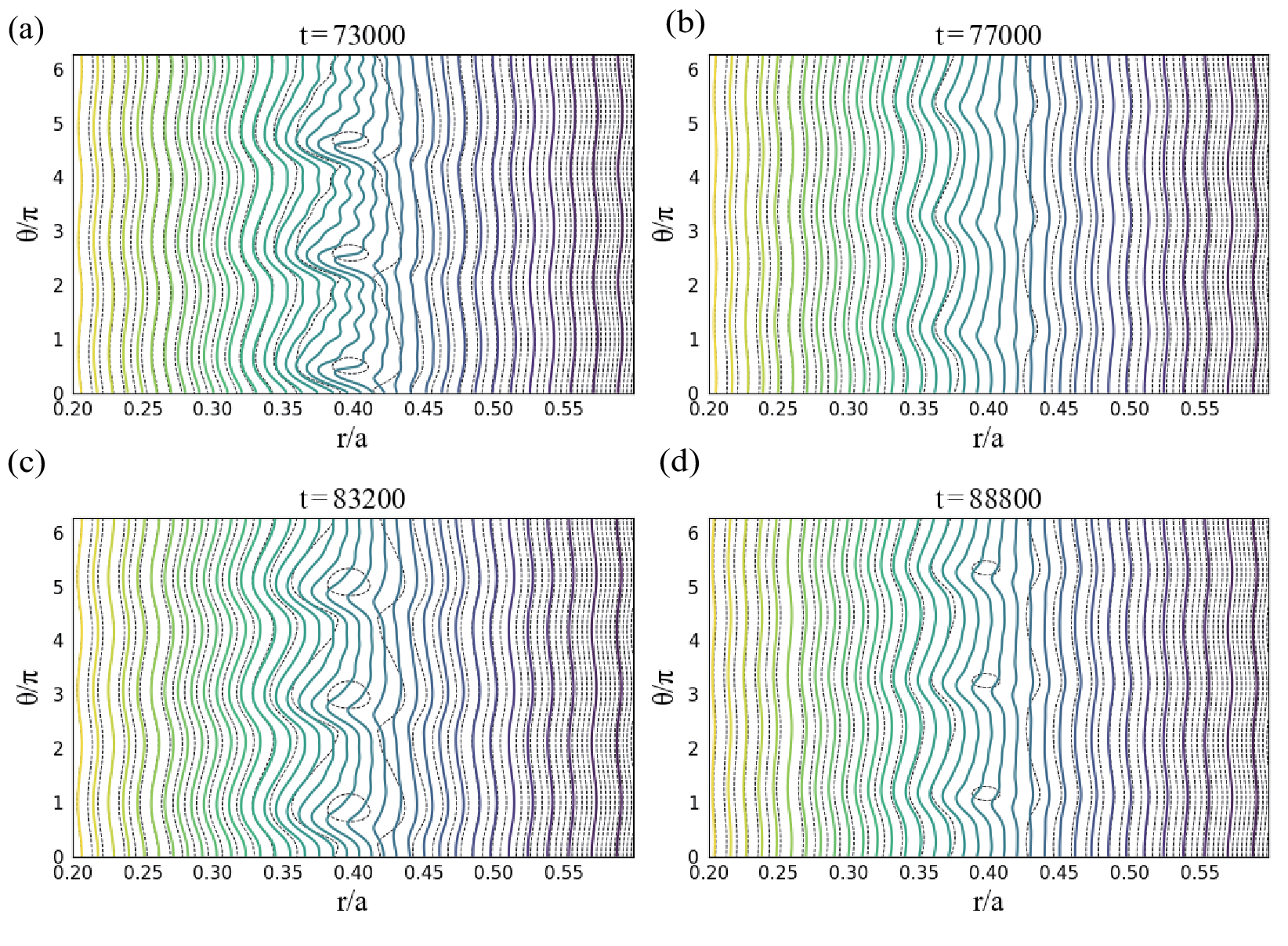}
\caption{Contour plot of the plasma pressure $p$ (solid lines) and poloidal flux $\psi$
(dotted lines), corresponding to the four red time points marked in figure \ref{fig9}.}
\label{fig10}
\end{figure*}

\begin{figure*}
\centering
\includegraphics[width=1.\textwidth]{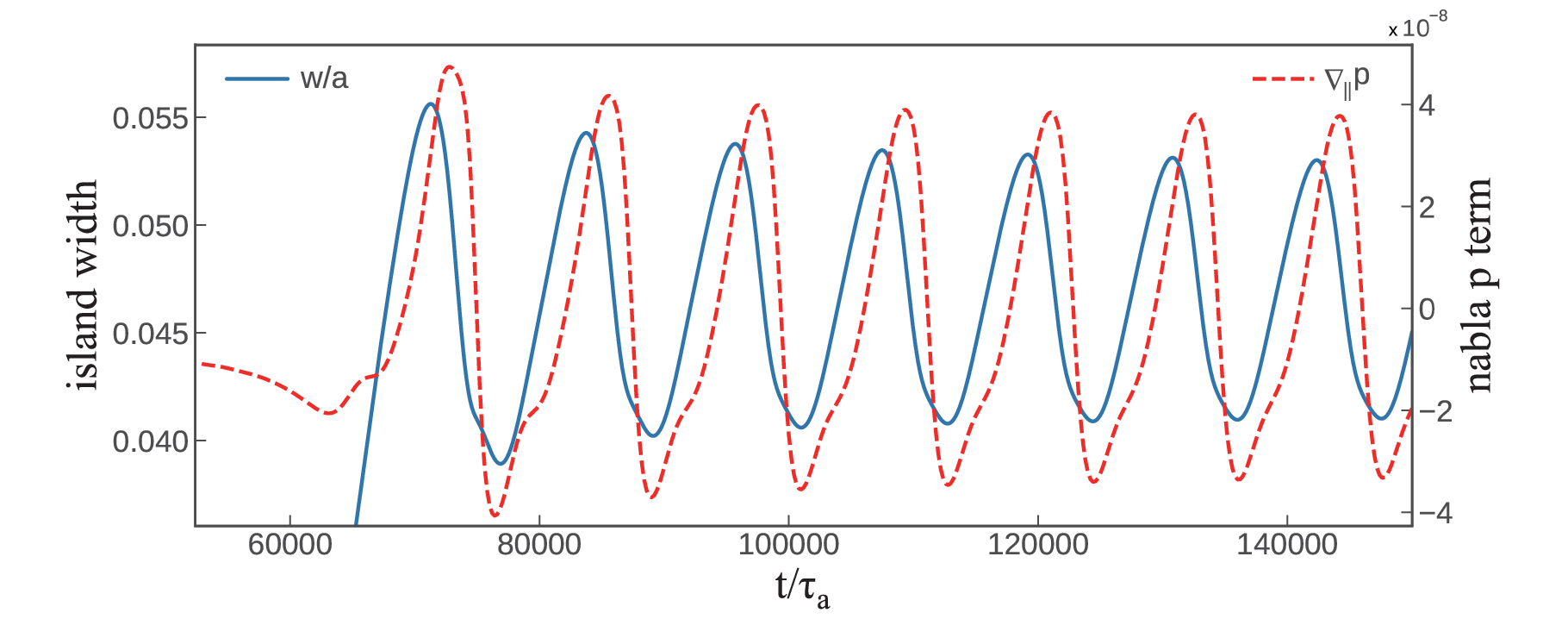}
\caption{Nonlinear evolution of the island width (solid) and the value of  $\nabla_\parallel p$ (dotted) at the resonant surface after mode penetration.}
\label{fig11}
\end{figure*}

When it comes to a larger rotation frequency, the situation is somewhat different.
Similar to figure \ref{fig6}, the nonlinear evolution of the island width for a larger natural frequency $\omega_0$=12.8 kHz
is presented in figure \ref{fig8}. It shows again that the penetration threshold is almost the same. However, there are two
differences observed. First, the saturated island width is evidently smaller for the case ($\omega_{\rm E0}$=0, $\omega_{\rm dia0}$=12.8 kHz)
than that of the case ($\omega_{\rm E0}$=12.8 kHz, $\omega_{\rm dia0}$=0), implying that the diamagnetic flow can drive a
stabilizing effect on the magnetic island. Second, an oscillation phenomenon of the magnetic island is discovered after
mode penetration. To further analyze this oscillation phenomenon, the island width and flow frequency of the oscillated
case in figure \ref{fig8} are plotted, as shown in figure \ref{fig9}. The periodical oscillation of diamagnetic flow frequency
is observed as well in figure \ref{fig9} (b). It should be noted that the oscillation of frequency lags behind the island width
a bit, suggesting that the change of island width results in the change of the diamagnetic flow frequency at first.
Since the diamagnetic flow is proportional to the pressure gradient, it can be straightforwardly inferred
that this phenomenon is related to the change of plasma pressure. To proceed a further step, the contour plot of the
pressure $p$ together with the poloidal magnetic flux $\psi$ is shown in figure \ref{fig10}, corresponding to the 
four red time points in figure \ref{fig9}. As the magnetic island grows larger ($t$=73000 and $t$=83200), the pressure gradient
$\nabla_\parallel p$ with respect to the magnetic field inside the island becomes larger. For a smaller island width
($t$=77000 and $t$=88800), on the contrary, the $\nabla_\parallel p$ is smaller. This modification of pressure gradient
can in return affect the island width by the $\delta\nabla_\parallel p$ term in equation (\ref{eq2}), leading to the above
oscillation phenomenon. To make it more clear, the nonlinear evolutions of the island width and the value of  $\nabla_\parallel p$ at the resonant surface are plotted in figure \ref{fig11}.  It is obviously shown that the island width and pressure gradient exhibit a negative feedback relationship, as discussed above. 

\begin{figure*}
\centering
\includegraphics[width=1.\textwidth]{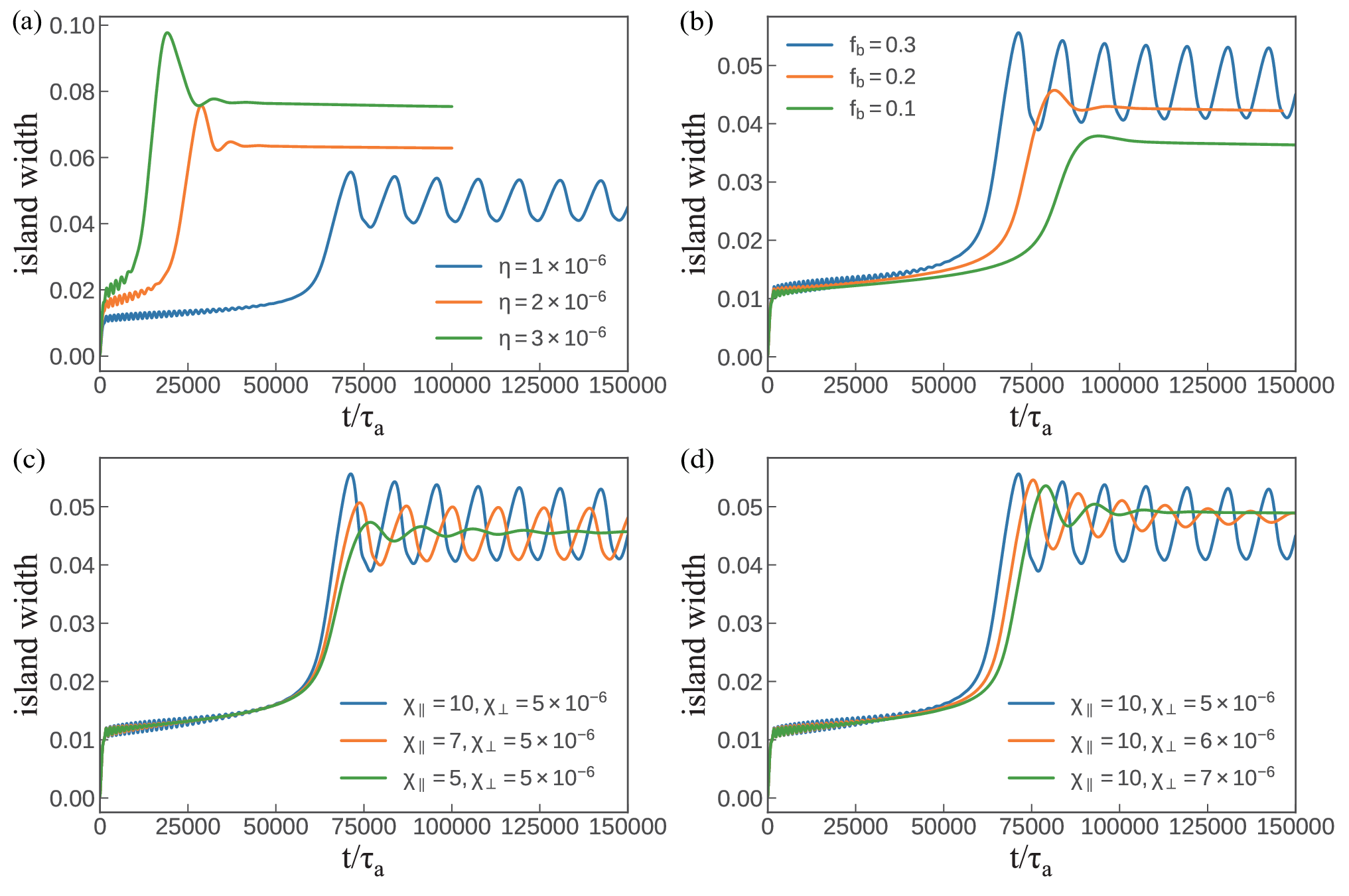}
\caption{Comparison of island width versus time for different (a) resistivity
$\eta$, (b) bootstrap current fraction $f_{\rm b}$, (c) parallel transport
coefficient $\chi_\parallel$ and (d) perpendicular transport coefficient 
$\chi_\perp$.}
\label{fig12}
\end{figure*}

In order to further verify our conjecture, the effect of resistivity $\eta$ is then investigated. For a larger
$\eta$, it turns out that the oscillation phenomenon disappears and the island width recovers as illustrated in
figure \ref{fig12} (a). It can be easily understood through equation (\ref{eq2}). Since $\eta$ is a diffusive term
, the effect of $\delta\nabla_\parallel p$ term, stabilizing the island and causing the oscillation, can be diffused to 
some extent with the increasing $\eta$, in much the same way as viscosity $\nu$ in the vorticity equation stabilizing the oscillation of rotation. In other words,   
the oscillation phenomenon is a result of the competition of the two terms $\delta\nabla_\parallel p$ and $\eta(j-j_{\rm b})$.
For the same reason, a smaller bootstrap current fraction $f_{\rm b}$ can remove the oscillation by making the $\eta(j-j_{\rm b})$ term larger,
shown in figure \ref{fig12} (b). 
As the ratio of parallel to perpendicular transport coefficients $\chi_\parallel/\chi_\perp$ is crucial to the
process of pressure evolution, the effects of different $\chi_\parallel$ and $\chi_\perp$ values are studied.
In figure \ref{fig12} (c) and (d), the temporal evolution of the island width is plotted for different $\chi_\parallel$
and $\chi_\perp$. The results are intuitive, i.e. a smaller $\chi_\parallel$ or larger $\chi_\perp$ can
eliminate the oscillation. This is because a smaller $\chi_\parallel/\chi_\perp$ can lower the energy transport
level along the magnetic field lines, which would reduce the variation of $\delta\nabla_\parallel p$ term when the
size of magnetic island changes.    

\section{Summary and discussion}\label{sec4}
The initial value code MDC (MHD@Dalian Code) is upgraded with the capability of two-fluid
effects. On the basis of the well-known four-field equations\cite{hazeltine1985}, the bootstrap current,
parallel and perpendicular transport effects are additionally included. In this work,
the numerical simulation on the mode penetration is conducted based on the two-fluid
model. Main points can be summarized as follows.

\begin{enumerate}
\item The threshold of mode penetration at zero rotation is explored. It is found that
for the classical TM ($f_{\rm b}=0$), there is not a threshold for mode penetration. At this
circumstance, the behavior of magnetic island is dominated by driven reconnection, i.e. 
the saturated island width is positively related to the amplitude of RMP. For the NTM ($f_{\rm b}\neq0$),
on the other hand, a mode penetration like phenomenon is observed consisting of a driven
reconnection regime and a NTM regime. This phenomenon is different from the so-called mode
penetration, but can be mistakenly defined as mode penetration if not carefully distinguished.
It may provide a possible explanation for the finite mode penetration threshold at zero rotation
detected in experiments. The polarization drift, which is an important physics for NTM beside the pressure transport model, is not included in this work. Its role will be investigated in the future study. 
\item The effect of diamagnetic drift flow on the mode penetration is numerically studied.
For a smaller diamagnetic drift flow, numerical results show that its influence is almost the same
as the electric drift flow with comparable frequency. However, for a larger diamagnetic drift
flow, it can drive a stabilizing effect on the magnetic island through the $\delta \nabla_\parallel p$
term in equation (\ref{eq2}). Besides, an oscillation phenomenon of the island width is observed.
This oscillation is linked with the change of pressure during the variation of island width.
It tends to appear in the high Lundquist number $S$ and high $\chi_\parallel/\chi_\perp$ regime, 
where the parameter of advanced tokamak exactly lies in.   
\end{enumerate}
 
\section*{Acknowledgements}
We acknowledge the Super Computer Center of Dalian University of Technology for providing computer resources. This work is supported by the National Key R\&D Program of China (No.2022YFE03040001), National Natural Science Foundation of China (Nos. 11925501 and 12075048), Chinese Academy of Sciences, Key Laboratory of Geospace Environment, University of Science \& Technology of China (No. GE2019-01) and Fundamental Research Funds for the Central Universities (No. DUT21GJ204).

\end{CJK*}
\end{document}